\documentclass[aps,prb,manuscript]{revtex4}
\usepackage{graphicx}
\usepackage{epsfig}
\usepackage{color}
\usepackage{dcolumn}
\usepackage{bm}
\begin{document}

\title{Nanogroove array on thin metallic film as planar lens with tunable focusing}

\author{L. David Wellems$^{1}$, Danhong Huang$^{1,\,\ast}$, \underline{T. A. Leskova}$^{2}$ and A. A. Maradudin$^{2}$}
\affiliation{$^{1}$Air Force Research
Laboratory, Space Vehicles
Directorate, Kirtland Air Force Base, NM 87117 USA\\
$^{2}$Department of Physics and Astronomy and Institute for Surface and Interface Science\\
University of California, Irvine, CA 92697 USA}

\date{\today}

\begin{abstract}
Numerical results for the distributions of light transmitted through metallic planar
lenses composed of symmetric nanogroove arrays on the surfaces of a gold film are presented and explained.
Both the near- and far-field distributions of the
intensity of light transmitted are calculated by using a Green's function formalism.
Results for an optimal transverse focus based on a quadratic variation of groove width are obtained.
Meanwhile, a significant dependence of the focal length on the wavelength of light
incident from the air side through the gold film into a dielectric substrate is found for this detector
configuration.
\end{abstract}

\maketitle

\section{Introduction}
\label{sec1}

There is considerable interest at the present time in shaping the spatial dependence
of the intensity of light transmitted through a metal film pierced by a one-dimensional
array of slits with subwavelength widths, as an alternative to refractive lensing. Sun and
Kim\,\cite{sun} have studied numerically the transmission of light through a finite periodic array
of subwavelength slits piercing a free-standing metal film with a convex dependence of its
thickness. The incident beam is focused by this metallic lens. Shi {\em et al.}\,\cite{shi} studied numerically
the focusing of light transmitted through a metallic film of constant thickness pierced by
an array of equally spaced nanoslits of varying widths. It was argued that the focusing
action is based on propagation across the opaque metal film of surface plasmon polaritons
supported by the slits (metal-dielectric-metal structures) of varying widths, thus ensuring an
enhanced transmission and a phase change along the film surface that lead to the constructive
interference. The focusing of light by such a structure deposited on a dielectric substrate
was demonstrated experimentally by Verslegers {\em et al.}\,\cite{fan}. The experimental results were in
an excellent agreement with FDTD simulations\,\cite{fan}. However, the focus distance was much
shorter than predicted by the model proposed previously.
\medskip

In this paper we demonstrate the focusing effect by calculating numerically the spatial
distribution of the intensity of light transmitted through a metal film sandwiched between
a cladding and a dielectric substrate whose surfaces are modeled by two finite aligned and
reversed arrays of nanogrooves of finite depth.
We have chosen to define transmissivity as the squared modulus of the ratio
of the transmitted to the incident {\bf H}-field amplitude for $p$ polarization.
In this case there are no surface polaritons
propagating through the film as there are no slits that completely pierce the metallic
film. In the case of the extraordinary transmission of light through a metal film with a
periodic nanoslit array, the surface polaritons supported by the slits play a minor role in
the phenomenon, and what is more it is not necessary to have slits that completely pierce
the metallic film to achieve the enhanced transmission\,\cite{maradudin1}. When a periodic nanogroove array
is illuminated by $p-$polarized light, whose magnetic vector is parallel to the generators
of the array, surface plasmon polaritons (SPPs) associated with the film-cladding and the
film-substrate interfaces are excited\,\cite{maradudin1,wellems}, and are diffracted by the structure into transmitted
volume waves in the substrate. The periodicity of the array enhances the excitation of surface
plasmon polaritons associated with the cladding-film interface and the conversion of
surface plasmon polaritons associated with the substrate-film interface into volume waves in
a range of SPP frequencies whose wavenumbers are in the vicinity of the boundary of the
second Brillouin zone, i.e. when $\lambda_{sp}\sim d$, where $d$ is the period of the array and $\lambda_{sp}$ is the
surface plasmon polariton wavelength. As a result the transmission through gold and silver
films as a function of wavelength\,\cite{maradudin1} shows sharp peaks and dips in a range of SPP frequencies
whose wavenumbers are near the boundary of the second Brillouin zone at both interfaces
of the film.
\medskip

It is also not necessary to have slits that completely pierce the metallic film to achieve
the focusing of light transmitted through it when the array is finite. The conversion of
either surface polaritons supported by the slits or surface polaritons supported by the film
itself into the volume wave in the substrate is a diffraction process, and since the system
is two-dimensional it produces volume cylindrical waves propagating away from the surface
into the substrate. A finite number of secondary sources of cylindrical waves separated
by subwavelength distances can produce the field focusing effect. The additional phases of
each of the secondary sources changes the focal depth and width, but affect only slightly
the focus distance. The effect can be modeled as follows. The periodic array of slits or
nanogrooves ensures the transmission through the otherwise nontransparent film, while the
now transparent aperture produces the focusing, in exactly the same manner as a refractive
nanolens in a diffractive regime. The focus distance is then determined primarily by the
aperture size, i.e. by the size of the array, and the wavelength of light in the substrate. The
intensity distribution in the far field is then determined by the Fresnel diffraction by the
aperture.
\medskip

In this paper we discuss focused patterns of nanogroove arrays and do not elaborate on
differences between nanoslit and nanogroove arrays, since they are physically equivalent in
producing the focusing effect except for some quantitative difference in the intensity of the
transmitted field. A simulation study was undertaken to search for the best way to improve
the quality of focus by changing the groove profile and groove width variation. Planar
nanolenses will have numerous applications in polarimetric imaging devices, solar cells, light
emitting diodes, and nanophotonics systems. Some preliminary results of this work was reported earlier.\,\cite{huang2}
\medskip

This paper is organized as follows. In Section\,II, we employ a previously developed
model and formalism,\,\cite{maradudin1,wellems} modified to include an arbitrary sequence of groove width variation.
Based on this formalism, numerical results are presented for the comparison of the focused
patterns of field intensity produced by different sequences of groove width variation and
different groove shapes. The conclusions drawn from these results are briefly summarized in
Section\,III.

\section{Model and Numerical Results}
\label{sec2}

\begin{figure}[htbp]
\begin{center}
\includegraphics[width=\columnwidth]{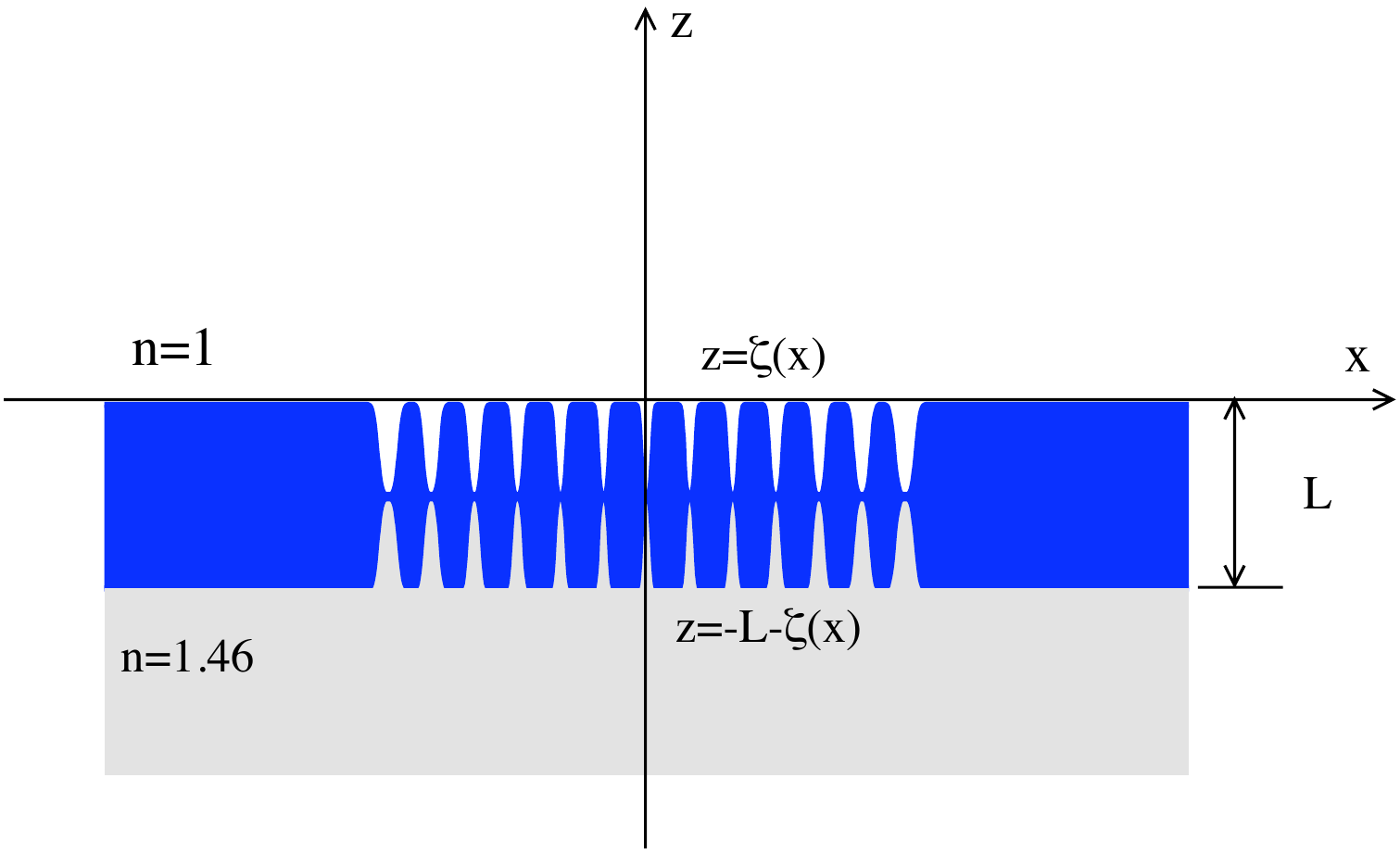}
\end{center}
\caption{(Color online) The structure studied, where $n$ is the index of refraction, $x,\,z$ are the spatial directions,
$\xi(x)$ is the surface profile function, and $L$ is the thickness of the unpatterned gold film.}
\label{f0}
\end{figure}

In this paper, we consider the model shown in Fig.\,\ref{f0} for a nanogroove array, in which the top [$\xi_1(x)=\xi(x)$ at $z=0$] and bottom [$\xi_2(x)=-\xi(x)$ at $z=-L$] surface profile functions\,\cite{maradudin1} employed
for modeling a nanogroove array patterned in a thin gold film are chosen to be

\begin{equation}
\xi(x)=-t\sum_{j=-M}^{M}\,\exp\left[-\left(\frac{x-jd}{b_j}\right)^s\right]\ ,
\label{profile}
\end{equation}
where $L$ is the thickness of the unpatterned gold film, $j$ is the index for labeling grooves,
$2M+1$ is the total number of grooves in the array, $t<L/2$ is the groove depth, $d$ is the
period of the groove array, and $\{b_j\}$ is an arithmetic sequence representing a specific pattern
of groove-width variation. The surface profile functions $\xi(x)$ in Eq.\,(\ref{profile}) are symmetric with
respect to the middle groove centered at $x=0$, and $s=2$ (or $s=4$) corresponds to a
Gaussian (or a quartic) functional form for a groove, respectively. The arithmetic sequence
$\{b_j\}$ in Eq.\,(\ref{profile}) is assumed to be

\begin{equation}
b_j=\alpha+\beta\,|j|+\gamma\,|j|^2\ ,\ \ \ \ \ \ \ \ \mbox{for}\ |j|\leq M\ ,
\end{equation}
where $\alpha$ represents the width of the central groove, and $\beta=0$ (or $\gamma=0$) corresponds to a
quadratic (or a linear) groove-width variation, separately. The spatial distributions of the
electromagnetic fields on the air side ($n=n_a$), inside the patterned gold film ($n=n_m$), and on the side of the
dielectric substrate ($n=n_s$) can be calculated by using a Green's function formalism.\,\cite{maradudin1,wellems}
\medskip

In our numerical calculations, whose results are presented below, we assume that the
metal film is illuminated by a normally incident $p-$polarized plane wave, whose magnetic-field component
$H_y(x,\,z)$ has a unit amplitude, or by a normally incident $s-$polarized plane
wave, whose electric-field component $E_y(x,\,z)$ has a unit amplitude. The parameters defining
the film have the values $L=0.4\,\mu$m, $t=0.196\,\mu$m, $d=0.25\,\mu$m, $M=6$, and $\alpha=40$\,nm
for the width of the central groove. The frequency-dependent complex refractive index $n_m$
for the gold film is obtained by interpolation from the data in the paper by Johnson and
Christy\,\cite{christy}. The values of the other parameters, such as $n_{a,\,s}$, $\beta$, $\gamma$, $s$, and the incident light
wavelength $\lambda_0$, used in our numerical calculations will be given in the figure captions.

\subsection{Focusing by a finite aperture}

\begin{figure}[htbp]
\begin{center}
\includegraphics[width=\columnwidth]{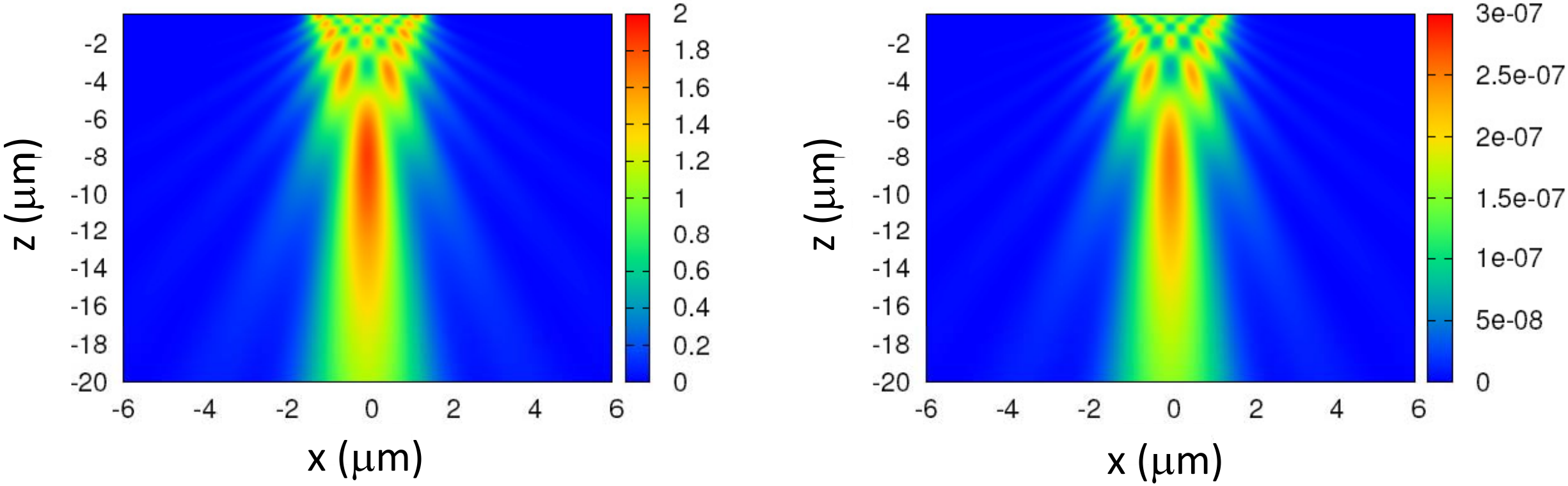}
\end{center}
\caption{(Color online) Color level plot of the spatial distribution of the intensity of the $p-$ (left panel)
and $s-$polarized (right panel) light of $\lambda_0=0.63\,\mu$m transmitted through the gold film. A quartic functional
form of the same width ($\beta=\gamma=0$) is used for the grooves constituting the surface profiles. The medium of
incidence is air $n_a=1$, while the substrate is a dielectric with the refractive index $n_s=1.46$.}
\label{f1}
\end{figure}

In Fig.\,\ref{f1} we present color level plots of the intensity of the field of light $|H_y(x,\,z)|^2$ (left) of $p$
and $|E_y(x,\,z)|^2$ (right) of $s$ polarization transmitted through the gold film from the air cladding
into the dielectric substrate. The surface profile functions are arrays of periodic grooves of
the same half widths ($\beta=\gamma=0$).
\medskip

As can be seen from plot presented in Fig.\,\ref{f1} the interference pattern of the transmitted
light of both polarizations exhibits a focal spot with the same focal distance but the intensity
of the $s-$polarized field is seven orders of magnitude weaker. The additional interference
maxima are of the same strength as the primary focal spot. Thus, in spite of the fact that
the groove arrays considered in our calculations are finite (the aperture size is $1.5\,\mu$m), and
the period is significantly smaller then the wavelength of light in air and in the substrate,
the enhanced transmission in $p-$polarization is obvious. The intensity distribution of the
primary focal spot is described by

\[
\left|H_y(x,\,z)\right|^2\sim\left|\left\{C\left[\frac{2n_s\,(x + {\cal D}/2)}{\sqrt{\lambda_0\,|z + L|}}\right]-C\left[\frac{2n_s\,(x - {\cal D}/2)}{\sqrt{\lambda_0\,|z + L|}}
\right]\right\}\right.
\]
\begin{equation}
-i\left.\left\{S\left[\frac{2n_s\,(x + {\cal D}/2)}{\sqrt{\lambda_0\,|z + L|}}\right]-S\left[\frac{2n_s\,(x -{\cal D}/2)}{\sqrt{\lambda_0\,|z + L|}}\right]\right\}\right|^2\ ,
\end{equation}
where $S(x)$ and $C(x)$ are the Fresnel integrals, and ${\cal D}$ is the array length. The focal length
(the position of the maximum of the intensity) is $z_0\sim n_s{\cal D}^211/(28\lambda_0)$, and is inversely
proportional to the wavelength of light in the substrate $\lambda_0/n_s$. From the calculated
$|H_y(x,\,z)|^2$ ($p$-polarization) and $|E_y(x,\,z)|^2$ ($s$-polarization) with groove width variation (not shown here), we find that
while the constructive interference is completely destroyed
by introducing the quadratic groove width variation in the case of $s-$polarized light, the
quality of the primary focal spot is considerably improved and the subsidiary maxima of
the interference pattern are suppressed in the case of $p-$polarized light. Note that the focal
length decreases. From the calculated total (integrated over the transmission
angles from $-\pi/2$ to $\pi/2$) transmission coefficient as a function of the wavelength of the
$p-$polarized incident light for different combinations of the cladding and substrate materials (not shown here),
we clearly see the enhanced transmission in comparison with $n_a=n_s=1$, although the array is limited to a few
wavelengths. In addition, the introduction of a weak quartic change in the width of the grooves leads
to a slight shift in the position and the broadening of the enhanced transmission peak, but does not destroy it.

\subsection{Wavelength effect on the focal length}

\begin{figure}[htbp]
\begin{center}
\includegraphics[width=\columnwidth]{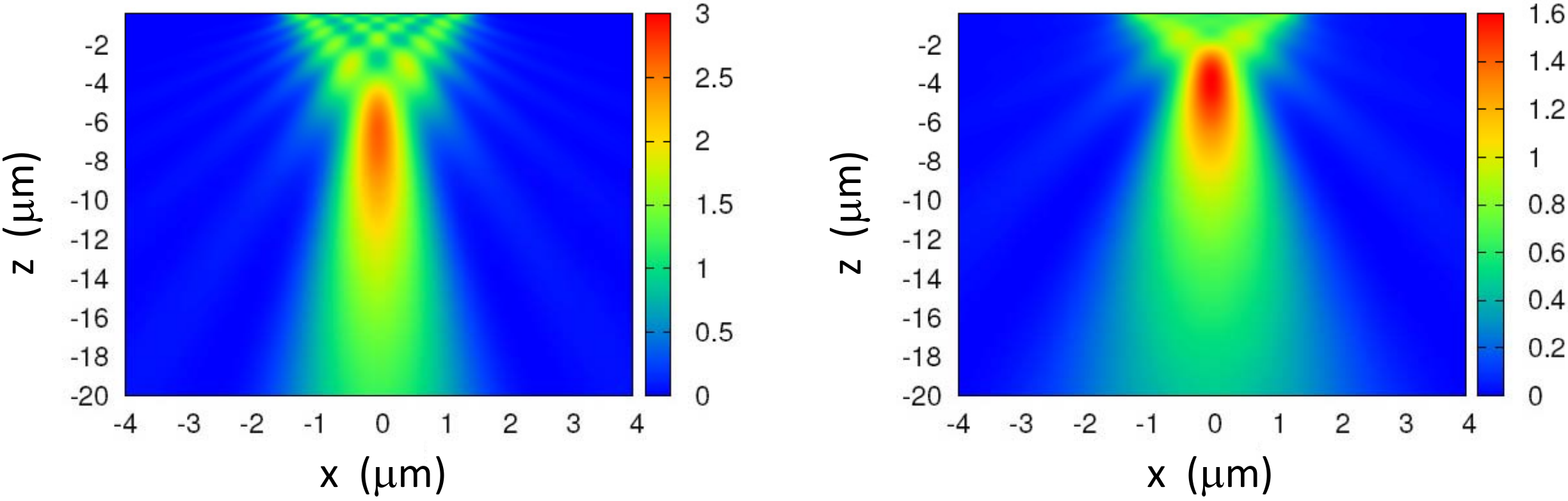}
\end{center}
\caption{(Color online) Contour plot of the spatial distribution of $|H_y(x,\,z)|^2$ for two incident-light
wavelengths, $\lambda_0=0.63\,\mu$m (left panel) and $\lambda_0=1\,\mu$m (right panel). Here, a quartic functional
form is used and the light is incident from the upper air side (detector configuration). The other
parameters in the calculations are $n_s=1.46$, as well as $\beta=0$ and
$\gamma=35/36$\,nm for the quadratic
(or parabolic) groove-width variation.}
\label{f2}
\end{figure}

Figure\ \ref{f2} presents a comparison of the spatial distributions of $|H_y(x,\,z)|^2$ when $p-$polarized
light is incident normally from the upper air side (corresponding to a detector configuration)
at $\lambda_0=0.63\,\mu$m (left panel) and $\lambda_0=1\,\mu$m (right panel). Here, a quartic functional form
($s=4$) is assumed for the symmetric nanogroove array on a gold film, which has a quadratic
groove-width variation ($\beta=0$). It is clear from Fig.\,\ref{f2} that the focal spot shifts upward from
$z=-7\,\mu$m to $z=-5\,\mu$m when $\lambda_0$ is increased from $0.63\,\mu$m to $1\,\mu$m. The positions of the
maxima of the intensity are in agreement with the inverse dependence of the focal length
on the wavelength. Meanwhile, the longitudinal ($z$ direction) size of the focal spot shrinks,
although its transverse ($x$ direction) size remains constant. This provides a possibility for
multi-color detection if several active detection layers with specific absorption wavelengths
are embedded at different depths underneath the nanogroove array on top of a quantum-well
photodetector.\,\cite{huang1} The magnification also decreases, which is obviously related to the
fact that the overall transmission decreases as the wavelength increases.
\medskip

\begin{figure}[htbp]
\begin{center}
\includegraphics[width=\columnwidth]{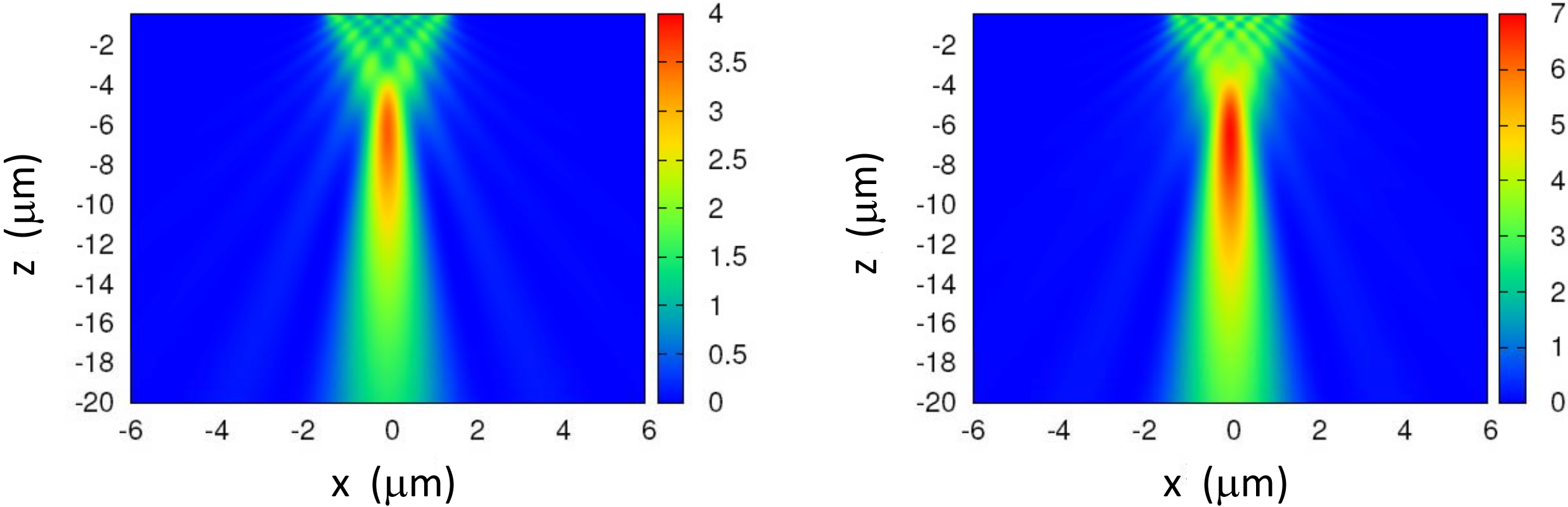}
\end{center}
\caption{(Color online) Contour plot of the spatial distribution of $|H_y(x,\,z)|^2$ for two incident-
light wavelengths and different substrates $\lambda_0=0.63\,\mu$m , $n_s=1.56$ (left panel) and $\lambda_0=1\,\mu$m,
$n_s=2.61$ (right panel). Here, a quartic functional form is used and the light is incident from the
upper air side (detector configuration). The other parameters in the calculations are $\beta=0$ and
$\gamma=35/36$\,nm for the quadratic groove-width variation.}
\label{f3}
\end{figure}

In Fig.\,\ref{f3} we present the color level plot of the spatial distribution of the field intensity
in the case when the dielectric substrate is adjusted to produce the strongest transmission
at the given wavelength. In this case, the wavenumbers of SPP polaritons associated with
the film-substrate interface are the same, so that at both wavelengths the effective aperture
for SPP is the same. This is the reason for the same focal distance. It is clearly seen that
the magnification is greatly increased for both chosen wavelengths.

\subsection{Difference in detector and emitter configurations}

\begin{figure}[htbp]
\begin{center}
\includegraphics[width=\columnwidth]{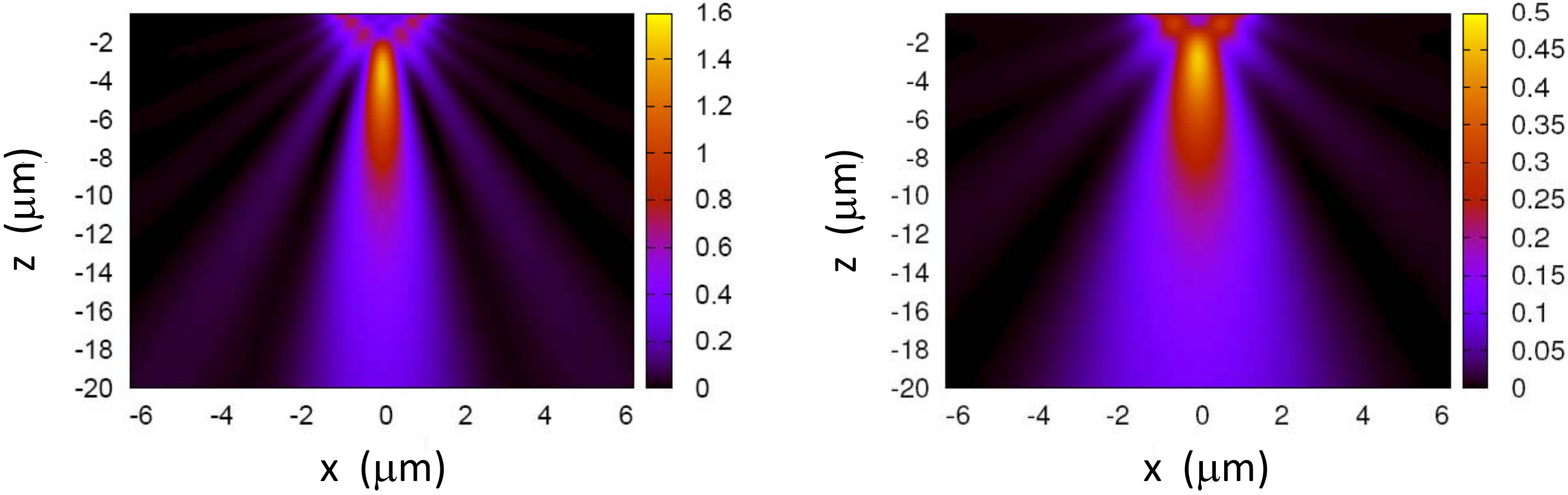}
\end{center}
\caption{(Color online) Contour plots of $|H_y(x,\,z)|^2$ at $\lambda_0=0.63\,\mu$m (left panel) and $\lambda_0=1\,\mu$m
(right panel). Here, a quartic functional form is assumed and the light is incident from the upper
substrate side (emitter configuration). The other parameters are $n_a=1.46$, $n_s=1$, as well as
$\beta=0$ and $\gamma=35/36$\,nm for the quadratic groove-width variation.}
\label{f4}
\end{figure}

For device-configuration comparison, we show in Fig.\,\ref{f4} the results for $|H_y(x,\,z)|^2$ when
the roles of air and dielectric substrate are interchanged, so that $p-$polarized light is incident
normally from the upper cladding side with $n_a=1.46$ and is transmitted into the air $n_s=1$
(corresponding to an emitter configuration) at $\lambda_0=0.63\,\mu$m (left panel) and $\lambda_0=1\,\mu$m
(right panel), respectively. Here, a quartic functional form for a groove, as well as a quadratic
groove-width variation, are also assumed for modeling the nanopatterned symmetric groove
array. In this case, however, we find no shift of the focal spot in the $z$ direction with a change
of $\lambda_0$. In this configuration the wavelength in the substrate (air) is considerably larger than in
the case described in the preceding section. This leads to shorter focal length ($1/1.46$ times).
Since the decay length $\zeta$ of the groove near field in the dielectric substrate side scales like $\zeta\sim\lambda_0/n_s$,\,\cite{raether}
the inverse dependence of the focal length on the wavelength will be completely masked
by the increase of $\zeta$ with $\lambda_0$. At the same time, the transmitted field is reduced, as is seen by comparing
Fig.\,\ref{f4} with Fig.\,\ref{f2}. In addition, the transverse size of the focal spot is enlarged on the air
side as $\lambda_0$ is increased from $0.63\,\mu$m to $1\,\mu$m. From the results shown in Fig.\,\ref{f4}, we expect
that this emitter configuration can be employed for constructing a wavelength-insensitive
planar lens with a very thin dielectric substrate that is transparent to light of wavelength
$\lambda_0\geq 0.63\,\mu$m.
\medskip

We also study the effect of a dielectric substrate on focusing
action from the calculated $|H_y(x,\,z)|^2$ for $p-$polarized light incident
normally from the upper air side at $\lambda_0=1\,\mu$m (not shown here), from which we find
that the focusing power almost disappears when the dielectric substrate is
replaced by air. Meanwhile, the non-focused bright spot is seen to move upward
as $n_s$ is reduced from $1.46$ to $1.0$. On the
other hand, the focal spot shifts downward as $n_s$ is increased from $1.0$ to $2.0$, in
agreement with the diffractive regime of the lens, which is accompanied by an expansion of
the longitudinal size of the focus spot.

\subsection{Effects of groove shape and groove-width variation}

\begin{figure}[htbp]
\begin{center}
\includegraphics[width=\columnwidth]{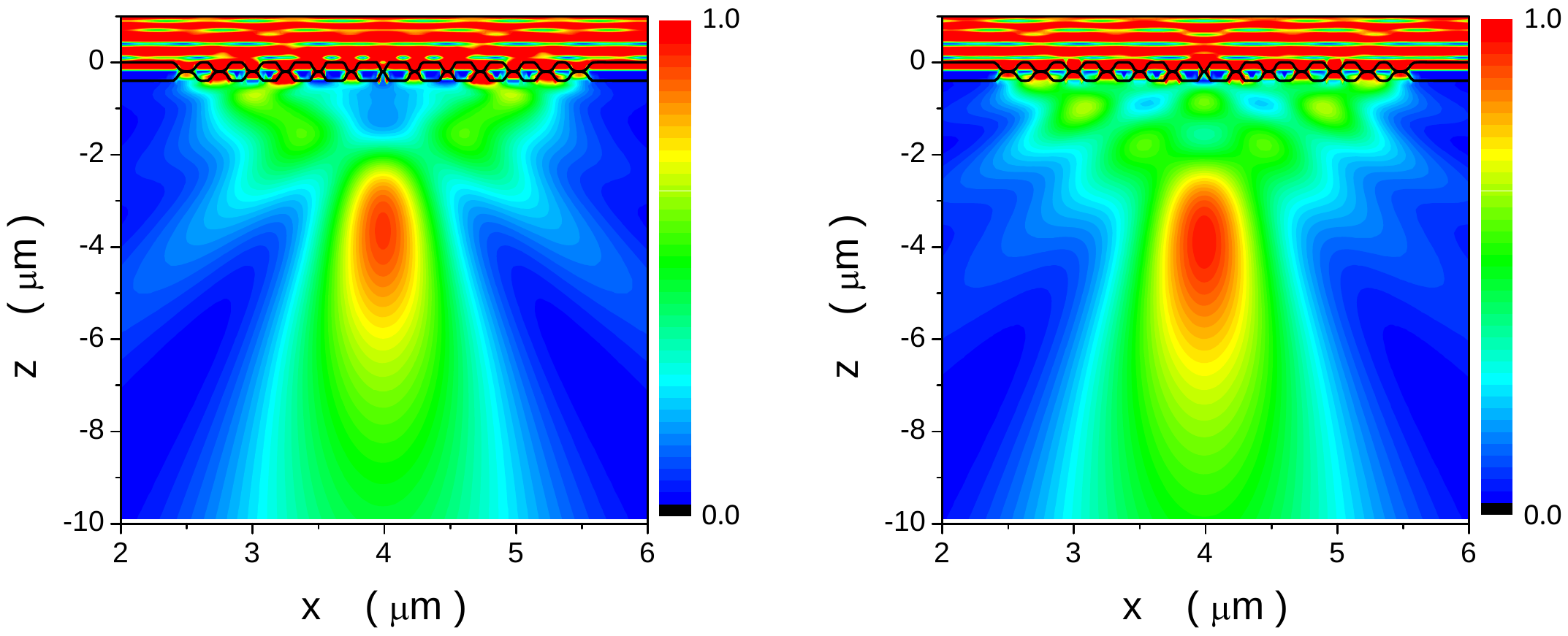}
\includegraphics[width=\columnwidth]{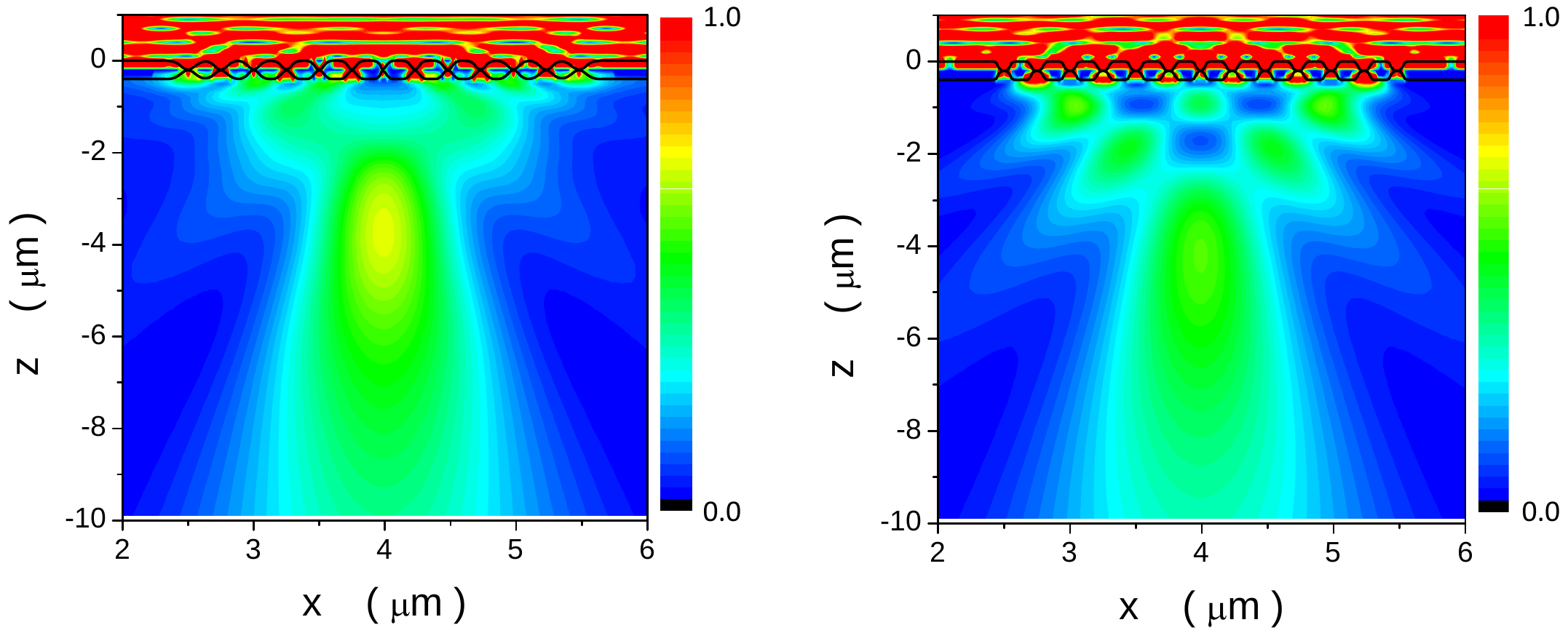}
\end{center}
\caption{(Color online) Contour plots of $|H_y(x,\,z)|^2$ with the quadratic groove-width variation
(upper-left panel) and the linear groove-width variation (upper-right panel). Here, a quartic
functional form is assumed for the upper two panels in this figure. We take $\beta=0$ and
$\gamma=35/36$\,nm for the quadratic groove-width variation and $\beta=35/6$\,nm and
$\gamma=0$ for the linear groove-width variation, respectively.
We also display here the contour plots for the scaled $|H_y(x,\,z)|^2$ with the
quadratic groove-width variation (lower-left panel) and with a constant groove width (lower-right
panel). Here, a Gaussian functional form is used for the lower-left panel, while a quartic functional
form is assumed for the lower-right panel. In addition, we take $\beta=0$ and $\gamma=35/36$\,nm for the
quadratic groove-width variation and $\beta=\gamma=0$ for the constant groove width, separately. The
light is incident from the upper substrate side at $\lambda_0=0.8\,\mu$m, and the refractive index of the
substrate is $n_s=1.46$, where $n_a=1$ and the black curves close to $z=0$ indicate the surface profiles of the patterned gold film.}
\label{f5}
\end{figure}

To study the effect of groove shape, as well as the effect of groove-width variation, we com-
bine four calculated spatial distributions for $|H_y(x,\,z)|^2$ in Fig.\,\ref{f5} obtained with $p-$polarized
light of wavelength $\lambda_0=0.8\,\mu$m incident normally in an emitter configuration. A quartic
functional form is assumed for the groove array (two upper panels) with a quadratic (left)
and a linear (right) groove-width variation, to investigate the effect of groove-width
variation. We also show the result for $|H_y(x,\,z)|^2$ (lower-right panel) with a constant groove
width. In addition, the calculated $|H_y(x,\,z)|^2$ (lower-left panel) for the case with a Gaussian
functional form ($s=2$) is included in Fig.\,\ref{f5} to demonstrate the groove-shape effect. By
comparing the two upper panels of Fig.\,\ref{f5}, we find that the quadratic variation of groove
width leads to an enhanced focusing power with a smaller spot size in both directions. For
a constant groove width, the focusing action is completely lost. As the corners of a groove
are rounded by a Gaussian functional form, in comparison with a quartic one, the focusing
power is partially suppressed, but the focal spot does not move at all.

\section{Conclusions and Remarks}
\label{sec3}

In conclusion, when light is incident from the upper air side, we have demonstrated that
the focal length of a planar metallic lens based on a variable nanogroove array deposited on
a dielectric substrate can be controlled by varying the wavelength of the incident light. How-
ever, this wavelength-tunability of the focal length for a metallic planar lens is completely
absent when the light is incident from the substrate side. These numerical results can be
applied to the design of a multi-color photodetector in which a number of active detection
layers with specific absorption wavelengths are embedded at different depths underneath
the top groove array. Moreover, the enhanced focusing power of a metallic planar lens with
a quadratic groove-width variation is observed in comparison with that of a linear one. A
range for the refractive index of a dielectric substrate is found as an imposed restriction
for the tunable focal length of a metallic planar lens. The sharpness of a groove corner is
shown to play an important role in the focusing power of a metallic planar lens through the
comparison of a groove in a quartic functional form with that in a Gaussian one.
\medskip

When a $p$-polarized incident light illuminates a periodic nanogroove array, SPPs, which are
associated with the film-cladding and film-substrate interfaces, can be excited. These excited
SPP waves will be diffracted by the array into volume cylindrical waves after their transmission as
long as the SPP wavenumbers fall into the vicinity of the boundary of the second Brillouin zone.
Diffraction induces a coupling between SPP modes with different reciprocal lattice vectors and
produces a gap at either the center or the boundary of the first Brillouin zone. These generated
cylindrical waves, which are separated by subwavelength distances, lead to the field focusing effect.
We note that the periodic array of nanogrooves is only responsible for the transmission through
the otherwise nontransparent film. It is the transparent aperture that produces the focusing. As
a result, the focus distance is decided by both the size of the array and the wavelength of light in
the substrate, which allows for a tunable focusing.
\medskip

When the total number, 2M + 1, of grooves in a metallic planar lens is increased, we
expect to see a significant reduction in the transverse size of a focus spot, which facilitates
an even smaller pixel size for a photodetector focal-plane-array to improve its detectivity,
as well as a suppression of cross-talk between different pixels.\,\cite{levine}

\begin{acknowledgements}
We would like to thank the Air Force Office of Scientific Research (AFOSR) for its support.
\end{acknowledgements}

\newpage

\newpage

\newpage

\newpage

\newpage

\end{document}